\documentstyle[11pt]{article}
\def\cite#1{#1}
\newcommand{\ct}[1]{[\cite{#1}]}
\def\thebibliography#1{\section*{References}\list
 {[\arabic{enumi}]}{\settowidth\labelwidth{[#1]}\leftmargin\labelwidth
 \advance\leftmargin\labelsep
 \usecounter{enumi}}
 \def\newblock{\hskip .11em plus .33em minus -.07em}
 \sloppy
 \sfcode`\.=1000\relax}

\begin{document}
\begin{center}
{\large\bf Asymptotic conditions for electromagnetic form factors of hadrons represented
by VMD model}
\vspace{1mm}
\end{center}
\vspace{0.1cm}
\begin{center}

Cyril Adamu\v s\v cin, Anna-Zuzana Dubni\v ckov\'a\\
{\em Dept. of Theoretical Physics, Comenius University, Mlynsk\'a dolina, 842 48 Bratislava, Slovak Republic}\\

Stanislav Dubni\v cka, Roman Pek\'arik\\
{\em Institute of Physics, Slovak Academy of Sciences,
D\'ubravsk\'a cesta 9, 842 28 Bratislava, Slovak Republic.}\\

and\\

Peter Weisenpacher\\
{\em Institute of Informatics, Slovak Academy of Sciences,
D\'ubravsk\'a cesta 9, 842 37 Bratislava, Slovak Republic.}\\

\vspace{0.1cm}
\end{center}

\begin{abstract}

A system of linear homogeneous algebraic equations for coupling
constant ratios of vector-mesons to hadrons is derived by imposing
an assumed asymptotic behaviour upon the VMD pole parametrization
of an hadron electromagnetic form factor. A similar system of
equations with a simpler structure of coefficients, taken as even
powers of vector-meson masses, is derived by means of integral
superconvergent sum rules for the imaginary part of the considered
form factor using its appropriate $\delta$-function approximation.
Although both systems have been derived starting from different
properties of the electromagnetic form factor and they look each
in its own way, it is shown explicitly that they are fully
equivalent.
\end{abstract}

\section{Introduction}

In a construction of the unitary and analytic model \ct{1-6} of
the electromagnetic (EM) structure of an arbitrary hadron, the
vector-meson-dominant (VMD) pole parametrization of EM form
factors (FF's)
\begin{equation}
F_h(t) = \sum_{v=1}^n\frac{m_v^2}{m_v^2-t}(f_{vhh}/f_v) \label{z1}
\end{equation}
is the starting point. The FF in this form is a pure real function
in the whole physical region $-\infty < t < +\infty$, where
$t=-Q^2$ is the squared momentum transfer of the virtual photon,
$m_v$ are the masses of vector mesons and $f_{vhh}$ and $f_v$ are
the coupling constants of vector-meson to hadron and
vector-meson-photon transition, respectively. The expression
(\ref{z1}) provides the basis for a consistent unification of all
other FF properties, like asymptotics, analytic properties,
unitarity, normalization etc, into an advanced unitary and
analytic model.

Based on the so-called quark counting rules (QCR), the asymptotic
behaviour of EM FF of a hadron is directly related \ct{7,8} to the
number of constituent quarks $n_q$ by the expression
\begin{equation}
F_h(t)_{|t|\to \infty} \sim t^{1-n_q}, \label{z2}
\end{equation}
From  Eq.(\ref{z2}) one can see immediately that for hadrons with
$n_q > 2$ the asymptotic behaviour of their EM FF's is different
from the asymptotic behaviour of (\ref{z1}) and so, in the
construction of an unitary and analytic model it has to be
suitably adopted.

First, one transforms the VMD parametrization (\ref{z1}) into
common denominator. As a result a rational function with a
polynomial of $(n-1)$ degree in the numerator is obtained. Then a
required asymptotic behaviour $\sim t^{-m} (m\le n)$ of the
considered FF is achieved by setting $(m-1)$ coefficients of the
highest powers of the variable $t$ in the numerator to be zero.

In this manner the first system of $(m-1)$ linear homogeneous algebraic equations for
coupling constant ratios with coefficients to be rather complicated sums of products of vector-meson
masses squared is formulated.

On the other hand, let us assume that FF is analytic in the whole
complex plane of its variable besides the cut on the positive real
axis from $t_0$ to $\infty$ and possesses the asymptotic behaviour
$\sim 1/t^m$. Then one can apply the Cauchy theorem to the FF with
an integration path being a circle of the radius $R\to \infty$,
but avoiding the cut on the positive real axis. Because the
integral along the circle with the infinity radius is zero, only
integrals on the upper and lower boundary of the cut will
contribute. By using the reality condition of FF they lead to the
integral superconvergent sum rule for the imaginary part of FF.

The same procedure can be applied to the functions $t F_h(t)$,
$t^2 F_h(t)$,$\cdots$,$t^{m-2} F_h(t)$ which lead to another
$(m-2)$ superconvergent sum rules for $t Im F_h(t)$, $t^2 Im
F_h(t)$, $\cdots$,$t^{m-2} Im F_h(t)$, respectively. Then
approximating FF imaginary part by $\delta$- function
appropriately one obtains another system of $(m-1)$ linear
algebraic equations for coupling constant ratios with coefficients
that are simply even powers of the corresponding vector-meson
masses.

In this paper we demonstrate that both systems of $(m-1)$
equations are equivalent.

The paper is organized as follows. In the next section we derive
two systems of $(m-1)$ linear algebraic equations starting from
different properties of electromagnetic  FF of a hadron. Section 3
is devoted to an explicit proof of their equivalence. In the last
section we present conclusions and discussion.

\section{Algebraic equations for coupling constant ratios}

Generally, let us assume that the FF in (\ref{z1}) is saturated by $n$
different vector meson pole terms and it has the asymptotic behaviour
\begin{equation}
F_{h_{|t|\to\infty}} \sim t^{-m}, \label{z3}
\end{equation}
where $m\le n$.

Transforming the VMD pole representation (\ref{z1}) into a common
denominator one obtains FF in the form of a rational function with
a polynomial of $(m-1)$ degree
\begin{equation}
P_{n-1}(t)= A_0 + A_1\cdot t + A_2\cdot t^2+\cdots + A_{n-1}\cdot t^{n-1} \label{z4}
\end{equation}
in the numerator, where
\begin{eqnarray*}
A_{n-1}&=&(-1)^{n-1}\sum_{j=1}^nm_j^2a_j \nonumber \\
A_{n-2}&=&(-1)^{n-2}\sum_{i=1\atop i\ne j}^n m_i^2\sum_{j=1}^n m_j^2a_j \nonumber \\
A_{n-3}&=&(-1)^{n-3}\sum_{i_1,i_2=1\atop i_1< i_2,i_r\ne j}^n m_{i_1}^2m_{i_2}^2\sum_{j=1}^n m_j^2a_j \nonumber \\
A_{n-4}&=&(-1)^{n-4}\sum_{i_1,i_2,i_3=1\atop i_1< i_2<i_3,i_r\ne j}^n m_{i_1}^2m_{i_2}^2m_{i_3}^2\sum_{j=1}^n m_j^2a_j \nonumber
\end{eqnarray*}
\begin{center}
\ldots \ldots \ldots
\end{center}
\begin{eqnarray}
A_{n-(m-1)}&=&(-1)^{n-m+1}\sum_{i_1,i_2,\cdots i_{m-2}=1\atop i_1< i_2\cdots <i_{m-2},i_r\ne j}^n m_{i_1}^2m_{i_2}^2\cdots m_{i_{m-2}}^2\sum_{j=1}^n m_j^2a_j \label{z5}\\
A_{n-m}&=&(-1)^{n-m}\sum_{i_1,i_2,\cdots i_{m-1}=1\atop i_1< i_2\cdots <i_{m-1},i_r\ne j}^n m_{i_1}^2m_{i_2}^2\cdots m_{i_{m-2}}^2m_{i_{m-1}}^2\sum_{j=1}^n m_j^2a_j \nonumber
\end{eqnarray}
\begin{center}
\ldots\ldots\ldots
\end{center}
\begin{eqnarray*}
A_2&=&(-1)^2\sum_{i_1,i_2,\cdots i_{n-3}=1\atop i_1< i_2\cdots <i_{n-3},i_r\ne j}^n m_{i_1}^2m_{i_2}^2\cdots m_{i_{n-3}}^2\sum_{j=1}^n m_j^2a_j   \\
A_1&=&(-1)\sum_{i_1,i_2,\cdots i_{n-2}=1\atop i_1< i_2\cdots <i_{n-2},i_r\ne j}^n m_{i_1}^2m_{i_2}^2\cdots m_{i_{n-3}}^2m_{i_{n-2}}^2\sum_{j=1}^n m_j^2a_j     \\
A_0&=&\sum_{i_1,i_2,\cdots i_{n-1}=1\atop i_1< i_2\cdots <i_{n-1},i_r\ne j}^n m_{i_1}^2m_{i_2}^2\cdots m_{i_{n-2}}^2m_{i_{n-1}}^2\sum_{j=1}^n m_j^2a_j
\end{eqnarray*}
and $a_j=(f_{jhh}/f_j)$.

In order to achieve the assumed asymptotic behaviour (\ref{z3})
one requires in (\ref{z4}) the first $(m-1)$ coefficients from the
highest powers of $t$ to be zero and as a result the following
first system of linear homogeneous algebraic equations for the
coupling constant ratios is obtained
\begin{eqnarray}
& &\sum_{j=1}^nm_j^2a_j=0 \nonumber \\
& &\sum_{i=1\atop i\ne j}^nm_i^2\sum_{j=1}^nm_j^2a_j=0\label{z6} \\
& &\sum_{i_1,i_2=1\atop i_1<i_2, i_r\ne j}^nm_{i_1}^2m_{i_2}^2\sum_{j=1}^nm_j^2a_j=0\nonumber \\
& &\sum_{i_1,i_2,i_3=1\atop i_1<i_2<i_3, i_r\ne j}^nm_{i_1}^2m_{i_2}^2m_{i_3}^2\sum_{j=1}^nm_j^2a_j=0\nonumber
\end{eqnarray}
\begin{center}
\ldots\ldots\ldots
\end{center}
$$\sum_{i_1,i_2,\cdots i_{m-2}=1\atop i_1<i_2\cdots <i_{m-2}, i_r\ne j}^nm_{i_1}^2m_{i_2}^2\cdots m_{i_{m-2}}^2\sum_{j=1}^nm_j^2a_j=0.$$

As one can see from (\ref{z6}) with increased $m$ the coefficients
become sums of more and more complicated products of squared
vector-meson masses.

For a derivation of the second system we employ the assumed
analytic properties of EM FF's of hadrons, consisting of infinite
number of branch points on the positive real axis, i.e. cuts. The
first cut extends from the lowest branch point $t_0$ to $+\infty$.
Then one can apply the Cauchy theorem to FF in $t$- plane
\begin{equation}
\frac{1}{2\pi i}\oint F_h(t)dt=0 \label{z7}
\end{equation}
where the closed integration path consists of the circle $C_R$ of
the radius $R\to \infty$ and the path avoiding the cut on the
positive real axis. As a result (\ref{z7}) can be rewritten into a
sum of the following four integrals
\begin{equation}
\frac{1}{2\pi i}\left\{\int_{C_R}F_h(t)dt+\int_{+\infty}^{t_0}F_h(t-i\epsilon)dt+ \int_{C_r/2}F_h(t)dt+\int^{+\infty}_{t_0}F_h(t+i\epsilon)dt \right\}=0
\label{z8}
\end{equation}
where $\epsilon \ll 1$ and $C_r/2$ is the half-circle joining the
upper boundary of the cut with the lower-boundary of the cut
around the lowest branch point $t_0$. The contribution of the
first integral in (\ref{z8}) is zero as $F_h(t)$ for $R\to \infty$
is vanishing. One can prove also that the third integral in
(\ref{z8}) for $r\to 0$ is zero. As a result one gets
\begin{equation}
\frac{1}{2\pi i}\int_{t_0}^{\infty}[F_h(t+i\epsilon)- F_h(t-i\epsilon)]dt=0.
\label{z9}
\end{equation}
Then, taking into account the reality condition of FF
\begin{equation}
F_h^*(t)= F_h(t^*) \label{z10}
\end{equation}
following from the general Schwarz reflection principle in the
theory of analytic functions, one arrives at the integral
superconvergent sum rule
\begin{equation}
\frac{1}{\pi}\int_{t_0}^{\infty} Im F_h(t)dt=0 \label{z11}
\end{equation}
for the imaginary part of the FF under consideration.

Repeating the same procedure for the functions $t F_h(t)$, $t^2
F_h(t)$, $\cdots$, $t^{m-2} F_h(t)$ which possess the same
analytic properties in the complex $t$-plane as $F_h(t)$, one gets
another $(m-2)$ superconvergent sum rules
\begin{eqnarray}
\frac{1}{\pi}\int_{t_0}^{\infty} t\cdot Im F_h(t)dt &=&0 \nonumber \\
\frac{1}{\pi}\int_{t_0}^{\infty} t^2\cdot Im F_h(t)dt &=&0 \label{z12}
\end{eqnarray}
$$
\ldots\ldots\ldots
$$
$$
\frac{1}{\pi}\int_{t_0}^{\infty} t^{m-2}\cdot Im F_h(t)dt =0.
$$
Now, approximating the FF imaginary part by $\delta$- function in
the following form
\begin{equation}
ImF(t)=\pi \sum_i^n a_i \delta(t-m_i^2)m_i^2 \label{z13}
\end{equation}
and substituting it into (\ref{z11}) and (\ref{z12}) one obtains the second
system of $(m-1)$ linear homogeneous algebraic equations for coupling constant ratios
$a_i=(f_{ihh}/f_i)$
\begin{eqnarray}
& & \sum_{i=1}^n m_i^2a_i =0 \nonumber \\
& & \sum_{i=1}^n m_i^4a_i =0 \nonumber \\
& & \sum_{i=1}^n m_i^6a_i =0 \label{z14}
\end{eqnarray}
$$
\ldots\ldots\ldots
$$
\begin{eqnarray*}
& & \sum_{i=1}^n m_i^{2(m-2)}a_i =0 \\
& & \sum_{i=1}^n m_i^{2(m-1)}a_i =0,
\end{eqnarray*}
where coefficients are simply even powers of the vector-meson masses.

In the next section we demonstrate explicitly that both systems of the
algebraic equations, (\ref{z6}) and (\ref{z14}), are equivalent.

\section{Equivalence of systems of algebraic equations for coupling constants ratios}

In this section we show step by step that the systems of linear
algebraic equations (\ref{z14}) and (\ref{z6}), despite the fact
that they have been derived starting from different properties of
the EM FF, and thus they appear to be different, are equivalent.
As a consequence, in a constructing the unitary and analytic model
of the EM structure of any hadron compound of more than two quarks
one can employ instead of equations (\ref{z6}) the simpler set
given by (\ref{z14}).

We start with the equations (\ref{z6}). From a direct comparison of
systems (\ref{z6})  and
(\ref{z14}) one can see immediately the identity of the first equations in
them.

The second equation in (\ref{z6}) can be written explicitly as follows
\begin{equation}
(m_2^2+m_3^2+\cdots +m_n^2)m_1^2a_1+ (m_1^2+m_3^2+\cdots
+m_n^2)m_2^2a_2+\cdots +(m_1^2+m_2^2+\cdots +m_{n-1}^2)m_n^2a_n=0.
\label{z15}
\end{equation}
Adding and substracting $m_1^4a_1$ to the first term of the sum, $m_2^4a_2$ to the
second term of the sum $\cdots$ etc. and finally $m_n^4a_n$ to the last term of the sum,
the equation  (\ref{z15}) can be modified into the form
\begin{equation}
\sum_{i=1}^nm_i^2\sum_{j=1}^nm_j^2a_j- \sum_{j=1}^nm_j^4a_j=0, \label{z16}
\end{equation}
from where one can see immediately that the second equation in
(\ref{z14}) is fulfilled
\begin{equation}
\sum_{j=1}^nm_j^4 a_j=0 \label{z17}
\end{equation}
as $\sum_{j=1}^nm_j^2 a_j=0$ is just the first equation in (\ref{z6})
and (\ref{z14}) as well.

The third equation in (\ref{z6})  can be written explicitly as follows
\begin{eqnarray}
& &(m^2_2m_3^2+ m^2_2m_4^2+\cdots + m^2_2m_n^2+ m^2_3m_4^2+ m^2_3m_5^2+\cdots  + m^2_3m_n^2+\cdots +m^2_{n-1}m_n^2)m_1^2a_1+ \nonumber \\
&+&(m^2_1m_3^2+ m^2_1m_4^2+\cdots + m^2_1m_n^2+ m^2_3m_4^2+ m^2_3m_5^2+\cdots  + m^2_3m_n^2+\cdots +m^2_{n-1}m_n^2)m_2^2a_2+\nonumber \\
&+&(m^2_1m_2^2+ m^2_1m_4^2+\cdots + m^2_1m_n^2+ m^2_2m_4^2+ m^2_2m_5^2+\cdots  + m^2_2m_n^2+\cdots +m^2_{n-1}m_n^2)m_3^2a_3+ \nonumber
\end{eqnarray}
$$
\ldots \ldots \ldots
$$
\begin{eqnarray}
&+&(m^2_1m_2^2+ m^2_1m_3^2+\cdots + m^2_1m_{n-2}^2+ m^2_1m_n^2+m^2_2m_3^2+\cdots  + m^2_2m_{n-2}^2+m^2_2m_n^2\cdots +\nonumber \\
&+& m^2_{n-2}m_n^2)m_{n-1}^2a_{n-1}+ \label{z18} \\
&+&(m^2_1m_2^2+ m^2_1m_3^2+\cdots + m^2_1m_{n-2}^2+ m^2_1m_{n-1}^2+ m^2_2m_3^2+\cdots  + m^2_2m_{n-2}^2+m^2_2m_{n-1}^2+\nonumber\\
&+&\cdots +m^2_{n-2}m_{n-1}^2)m_n^2a_n =0. \nonumber
\end{eqnarray}
Now adding and substracting all missing terms in (\ref{z18}) from
$\sum_{{i_1},{i_2}=1\atop {i_1}<{i_2}}^nm_{i_1}^2m_{i_2}^2 \sum_{j=1}^nm_j^2a_j$
which in the substraction form can be written explicitly as follows
\begin{eqnarray}
&-&(m_2^2+m_3^2+m_4^2+\cdots m_n^2)m_1^4a_1- \nonumber \\
&-&(m_1^2+m_3^2+m_4^2+\cdots m_n^2)m_2^4a_2- \nonumber \\
&-&(m_1^2+m_2^2+m_4^2+\cdots m_n^2)m_3^4a_3- \label{z19}
\end{eqnarray}
$$
\ldots \ldots \ldots
$$
\begin{eqnarray*}
&-&(m_1^2+m_2^2+\cdots +m_{n-2}^2+ m_n^2)m_{n-1}^4a_{n-1}-  \\
&-&(m_1^2+m_2^2+\cdots +m_{n-2}^2+ m_{n-1}^2)m_n^4a_n
\end{eqnarray*}
and again substracting and adding  $m_1^6a_1$ in the first line of
(\ref{z19}), $m_2^6a_2$ in the second line of (\ref{z19})...etc.,
and finally $m_n^6a_n$ in the last line of (\ref{z19}), one can
rewrite (\ref{z18}) into the form
\begin{equation}
\sum_{i_1,i_2=1\atop i_1<i_2}^n
m_{i_1}^2m_{i_2}^2\sum_{j=1}^nm_j^2a_j-\sum_{i=1}^nm_i^2\sum_{j=1}^nm_j^4a_j+\sum_{j=1}^nm_j^6a_j=0.
\label{z20}
\end{equation}
From this expression, taking into account the first two equations
in (\ref{z14}), the third equation of (\ref{z14})
\begin{equation}
\sum_j^n m_j^6a_j=0 \label{z21}
\end{equation}
follows.

The fourth equation in (\ref{z6}) takes the following explicit form
\begin{eqnarray}
& & (m^2_2m_3^2m_4^2+\cdots + m^2_2m_3^2m_n^2 + m^2_2m_4^2m_5^2+\cdots + m^2_2m_4^2m_n^2+\cdots  + \nonumber\\
&+&  m_{n-2}^2m_{n-1}^2m_n^2)m_1^2a_1+ \nonumber \\
&+& (m^2_1m_3^2m_4^2+\cdots + m^2_1m_3^2m_n^2 + m^2_1m_4^2m_5^2+\cdots + m^2_1m_4^2m_n^2+\cdots  + \nonumber\\
&+& m_{n-2}^2m_{n-1}^2m_n^2)m_2^2a_2+ \label{z22} \\
&+& (m^2_1m_2^2m_4^2+\cdots + m^2_1m_2^2m_n^2 + m^2_1m_4^2m_5^2+\cdots + m^2_1m_4^2m_n^2+\cdots  + \nonumber \\
&+& m_{n-2}^2m_{n-1}^2m_n^2)m_3^2a_3+ \nonumber
\end{eqnarray}
$$
\ldots \ldots\ldots
$$
\begin{eqnarray*}
 &+&(m^2_1m_2^2m_3^2+\cdots + m_1^2m_2^2m_n^2 + m_1^2m_3^2m_4^2+\cdots + m^2_1m_3^2m_n^2+\cdots + \\
&+&m_{n-3}^2m_{n-2}^2m_n^2)m_{n-1}^2a_{n-1}+  \\
&+&(m^2_1m_2^2m_3^2+\cdots + m^2_1m_2^2m_{n-1}^2 + m^2_1m_3^2m_4^2+\cdots + m^2_1m_3^2m_{n-1}^2+\cdots  +  \\
&+&m_{n-3}^2m_{n-2}^2m_{n-1}^2)m_2^na_n=0.
\end{eqnarray*}

First, adding and substracting all missing terms in (\ref{z22}) from
$$\sum_{{i_1},{i_2},{i_3}=1\atop i_1< i_2< i_3}^nm_{i_1}^2m_{i_2}^2 m_{i_3}^2\sum_{j=1}^nm_j^2a_j,$$
the equation (\ref{z22}) takes the form
\begin{eqnarray}
& &\sum_{i_1,i_2,i_3=1\atop i_1<i_2<i_3}^n m_{i_1}^2m_{i_2}^2m_{i_3}^2\sum_{j=1}^nm_j^2a_j- \nonumber \\
&-&\sum_{i_1,i_2=1\atop i_1<i_2, i_r \ne j}^nm_{i_1}^2m_{i_2}^2\sum_{j=1}^nm_j^4a_j=0. \label{z23}
\end{eqnarray}
Second, substracting and adding of all missing terms in (\ref{z23})
from
$\sum_{{i_1},{i_2}=1\atop i_1< i_2}^nm_{i_1}^2m_{i_2}^2 \sum_{j=1}^nm_j^4a_j$
one gets the equation
\begin{eqnarray}
& &\sum_{i_1,i_2,i_3=1\atop i_1<i_2<i_3}^n m_{i_1}^2m_{i_2}^2m_{i_3}^2\sum_{j=1}^nm_j^2a_j- \label{z24} \\
&-&\sum_{i_1,i_2=1\atop i_1<i_2}^nm_{i_1}^2m_{i_2}^2\sum_{j=1}^nm_j^4a_j+\sum_{i=1\atop i\ne j}^nm_{i}^2\sum_{j=1}^nm_j^6a_j=0. \nonumber
\end{eqnarray}
Finally, additions and substractions of all missing terms in (\ref{z24}) from
$\sum_{i=1}^nm_i^2\sum_{j=1}^nm_j^6a_j$ lead to the definitive form
of the fourth equation in (\ref{z6})
\begin{eqnarray}
& &\sum_{i_1,i_2,i_3=1\atop i_1<i_2<i_3}^n m_{i_1}^2m_{i_2}^2m_{i_3}^2\sum_{j=1}^nm_j^2a_j- \label{z25} \\
&-&\sum_{i_1,i_2=1\atop
i_1<i_2}^nm_{i_1}^2m_{i_2}^2\sum_{j=1}^nm_j^4a_j+\sum_{i=1}^n
m_i^2\sum_{j=1}^n m_j^6a_j-\sum_{j=1}^nm_j^8a_j=0. \nonumber
\end{eqnarray}
From here, taking into account the first three equations in
(\ref{z14}), the fourth equation in (\ref{z14})
\begin{equation}
\sum_{j=1}^nm_j^8a_j=0 \label{z26}
\end{equation}
follows.

It is now easy to give a straightforward generalization of the
above procedures:
\begin{itemize}
\item[i)] the q-th equation in (\ref{z6}) can be decomposed into q-terms
(see (\ref{z16}), (\ref{z20}) and (\ref{z25})) consisting of the
product of two parts, where the first part is just the sum of
decreasing numbers of products of different vector-meson masses
squared, starting from (q-1) coefficients and ending with the
constant 1. The second term takes the form
$\sum_{j=1}^nm_j^{\alpha}a_j$ with increasing even power $\alpha$
starting from $\alpha=2$ up to 2q.
\item[ii)]
there is an alternating sign in front of every term in that
decomposition, while the first term is always positive.
\end{itemize}

Now, in order to carry out a general proof of the equivalence of
the two systems of algebraic equations under consideration, let us
assume an equivalence of $(m-2)$ equations in (\ref{z6}) and
(\ref{z14}). Then, taking into account a generalization of our
procedure defined by rules $i)$ and $ii)$ above, one can decompose
the $(m-1)$-equation in (\ref{z6}) into the following form
\begin{eqnarray}
& & \sum_{i_1,i_2,i_3,\cdots ,i_{m-2} =1\atop i_1<i_2<i_3<\cdots <i_{m-2}}^n m_{i_1}^2m_{i_2}^2\cdots m_{i_{m-2}}^2\sum_j^n m_j^2a_j- \nonumber \\
&-&\sum_{i_1,i_2,i_3,\cdots ,i_{m-3} =1\atop i_1<i_2<i_3<\cdots <i_{m-3}}m_{i_1}^2m_{i_2}^2\cdots m_{i_{m-3}}^2\sum_{j=1}^nm_j^4a_j+ \label{z27} \\
&+& \sum_{i_1,i_2,i_3,\cdots ,i_{m-4} =1\atop i_1<i_2<i_3<\cdots <i_{m-4}}^n m_{i_1}^2m_{i_2}^2\cdots m_{i_{m-4}}^2\sum_j^n m_j^6a_j+\cdots +\nonumber \\
&+&(-1)^{m-3} \sum_{i=1}^nm_i^2\sum_{j=1}^nm_j^{2(m-2)}a_j+ (-1)^{m-2} \sum_{j=1}^nm_j^{2(m-1)}a_j=0,\nonumber
\end{eqnarray}
from where one can see immediately that the $(m-1)$ equation in
(\ref{z14}) is satisfied
\begin{equation}
\sum_{j=1}^nm_j^{2(m-1)}a_j=0 \label{z28}
\end{equation}
as  $\sum_{j=1}^nm_j^2a_j=0$,  $\sum_{j=1}^nm_j^4a_j=0$, $\cdots$,
$\sum_{j=1}^nm_j^{2(m-2)}a_j=0$ are just the first $(m-2)$ equations
in (\ref{z14}) assumed to be valid.

At the end we would like to draw an attention to the proof of the equivalence
of the systems of algebraic equations (\ref{z6}) and (\ref{z14})  from the
other point of view.

If the sums $\sum_{j=1}^nm_j^2a_j$,  $\sum_{j=1}^nm_j^4a_j$,
$\sum_{j=1}^nm_j^6a_j$, $\cdots$,  $\sum_{j=1}^nm_j^{2(m-3)}a_j$,
$\sum_{j=1}^nm_j^{2(m-2)}a_j$, $\sum_{j=1}^nm_j^{2(m-1)}a_j$ are
considered to be independent variables, then the first equation in
(\ref{z6}) together with the modified forms (\ref{z16}),
(\ref{z20}), (\ref{z25}),..,(\ref{z27}) form a system of $(m-1)$
homogeneous algebraic equations for these variables and the
equations (\ref{z14}) are just its trivial solutions.

\section{Conclusions and discussion}

Starting from different properties of the EM FF of hadron we have
derived two apparently distinct systems of linear homogeneous algebraic
equations for the coupling constant ratios of vector-mesons to the hadron
under consideration.

For a derivation of the first system of equations we have assumed
that EM FF of hadron is well approximated by a finite number of
vector-meson exchange tree Feynman diagrams leading to the VMD
pole parametrization of FF. A subsequent requirement of a true
asymptotics of EM FF gives the system of linear homogeneous
algebraic equations for coupling constant ratios with coefficients
that are rather complicated sums of products of squared
vector-meson masses.

For a derivation of the second system of equations analytic properties together
with the asymptotic behaviour of EM FF have been utilized. The application of Cauchy
theorem to  $F_h(t)$, $t F_h(t)$, $t^2 F_h(t)$, $\cdots$, $t^{m-2} F_h(t)$
leads to $(m-1)$ integral superconvergent sum rules for
$Im F_h(t)$, $t Im F_h(t)$, $t^2 Im F_h(t)$, $\cdots$, $t^{m-2} Im F_h(t)$.
Then an appropriate approximation of FF imaginary part by $\delta$ -function gives
another system of linear homogeneous algebraic equations for coupling constant
ratios with coefficients to be simply even powers of the vector-meson masses.

By using a sequence of algebraic manipulations it has been proved
step by step that both systems of equations for coupling constant
ratios are equivalent.

Finally a natural question arises about a practical application of
such systems of  equations for coupling constant ratios. The
latter was already indicated to some extent in the introduction,
but in relation to this issue some peculiarities have to be
mentioned. The asymptotic behaviour of EM FF is given by a number
$n_q$ of the constituent quarks in the hadron and so, the system
of algebraic equations for coupling constant ratios can be derived
only in the case if $n_q >2$. However, a necessary condition for
the latter is saturation of the sum in (\ref{z1}) by a number of
vector-meson resonances $n$ to be greater or equal to $n_q-1$ .

If $n>n_q-1$, then the derived system of  equations leads in a
constructed unitary and analytic model of EM structure of the
hadron to a remarkable reduction of a number of free coupling
constant ratios.

If $n=n_q-1$, then adding to $(n-1)$ algebraic equations the
equation following from a normalization  of $F_h(t)$ at $t=0$ that
typically has a non-zero value, one obtains inhomogeneous system
of $n$ linear algebraic equations with $n$ variables that can be
non-trivially solved for. Then solutions are just numerical values
of coupling constant ratios, which appear to be in very good
approximation to the physical reality (for the case of nucleons
see \ct9).

The inhomogeneous system of $n\ge n_q-1$ linear algebraic
equations is a very natural tool for a successful simultaneous
incorporation of the normalization and the true asymptotics of
$F_h(t)$ into the unitary and analytic model of EM structure of
any hadron with $n_q> 2$.

 The work was in part supported in  by Slovak Grant Agency for Sciences,
Grant No. 2/5085/20 (S.D.) and Grant No. 1/7068/20 (A.Z.D)

\end{document}